\documentclass[12pt,preprint,
aps,prl,nofootinbib,
showpacs,twoside,
superscriptaddress]{revtex4-2}

\usepackage{amsmath,bm}
\usepackage{graphicx}
\usepackage{amsfonts}
\usepackage{hyperref}
\usepackage[left=1.2in,right=1.2in,top=1.5in,bottom=1.5in]{geometry}
\usepackage{setspace}
\usepackage{orcidlink}
\usepackage{caption,subcaption}
\usepackage{ragged2e}
\newcommand{\red}[1]{\color{black}{#1}}

\renewcommand{\thefootnote}{*}

\newcommand{\hor}{\mathcal{H}}

\begin{document}

\setcounter{page}{0}

\begin{center}
    {\textbf{\Large{Loopy Black-Hole Remnants}}}\\[20pt]
    {\large Asier Alonso-Bardaji\footnote{asier.alonso@ehu.eus}}\\[4pt]
    \textit{\small Aix Marseille Univ., Univ. de Toulon, CNRS, CPT, UMR 7332, 13288 Marseille, France}
\end{center}

\vspace{16pt}

\noindent \textbf{Abstract}. The quantized area predicted by loop quantum gravity suggests the existence of a lower bound for black-hole horizons. We prove this intuition within a covariant effective model for spherical loop quantum gravity, where nonsingular quasi-static black holes evaporate until their horizons attain the smallest positive eigenvalue of the area operator. Consistent with the third law of black-hole thermodynamics, this final state---characterized by vanishing temperature and entropy---cannot be realized in finite time. The process thus leads to the formation of stable remnants, whose estimated masses are approximately 20.94$\mu$g, lying in the Planck regime.

\vspace{12pt}
 
\begin{flushright}
    {Submitted on March 12, 2025}
\end{flushright}

\vspace{1.5cm}
\vfill

\begin{center}
\noindent 
This essay received an Honorable Mention in the\\2025 Essay Competition of the Gravity Research Foundation.
\end{center}

\vfill

\renewcommand{\thefootnote}{\arabic{footnote}}

\newpage

At the final stages of black-hole evaporation, the temperature associated with the horizon diverges, and the once-massive object pops into a nakedly singular flat spacetime. This scenario is a direct---albeit likely unphysical---implication of quantum field theory applied to a classical curved spacetime background. Such an outcome is generally dismissed due to significant theoretical limitations, including strong backreaction effects, the breakdown of the quasistatic approximation, and anticipated quantum-gravitational corrections that could alter the process. A growing perspective posits the existence of metastable Planck-scale remnants, which may serve as candidates for dark matter~\cite{MacGibbon:1987my,Giddings:1992hh,Amadei:2021aqd} and potentially address the longstanding information-loss paradox. Interest in this hypothesis has been revitalized by the prospect of detecting such remnants~\cite{Perez:2023tld}, particularly within the framework of loop quantum gravity \cite{Bianchi:2018mml,Rovelli:2024sjl}. 
In fact, the pivotal prediction of a positive minimum eigenvalue for the area operator provides a key theoretical motivation for the existence of such remnants: the horizons of evaporating black holes cannot shrink smoothly to a point. 

As the full dynamics of loop quantum gravity continue to be unraveled, effective models have flourished in the literature. The basic variables in this theory are the holonomies of the Ashtekar-Barbero connection, evaluated along a closed path. This is effectively implemented  by replacing the connection components with sinusoidal functions, whose arguments include the so-called polymerization parameters. Their values are intrinsically linked to the loops of the holonomies and, thus, to the minimum area gap. In this way, they provide a bridge between the continuous regime and the underlying quantum geometry. 
{\red Different polymerization schemes have been the subject of recent attention, particularly in cosmological scenarios. Ranging from constant parameter choices to scale-dependent ones, each approach has its own advantages and drawbacks (see Sec.~IV~D of Ref.~\cite{Ashtekar:2018cay} for a detailed discussion).}
In the following, we review a recent black-hole model inspired by loop quantum gravity that {\red implements a generalized constant polymerization and} preserves the diffeomorphism invariance of general relativity~\cite{Alonso-Bardaji:2024tvp}. Later, we study its evaporation and show that it predicts the formation of zero-temperature and minimum-entropy Planckian remnants. 

Our focus is on the vacuum sector, which provides a single-parameter generalization of the Schwarzschild spacetime. It is described by the diagonal line element
\begin{align}\label{eq.metric}
    ds^2=&\,
    -\bigg(1-\frac{r_g}{r}\bigg)\bigg(1-\frac{r_0}{r}\bigg)^{-1}dt^2+\bigg(1-\frac{r_g}{r}\bigg)^{-1}\bigg(1-\frac{r_0}{r}\bigg)^{-2}dr^2 +r^2d\sigma^2,
\end{align}
with $d\sigma^2$ the metric of the unit two-sphere. The parameters $r_g$ and $r_0$ are positive length scales, related by $r_0/r_g=\lambda^2/(1+\lambda^2)$, where $\lambda\in\mathbb{R}^+$ is the polymerization parameter emerging from loop-quantum-gravity corrections \cite{Alonso-Bardaji:2024tvp,Alonso-Bardaji:2022ear}. This means that {\red the novel length scale $r_0$ is \textit{always} smaller than the radius of the black-hole horizon $r_g$}. However, let us momentarily set aside any notion of {quantumness} and study the above geometry with $0<r_0<r_g$.

\begin{figure}[t]
    \centering
    \includegraphics[width=.87\linewidth]{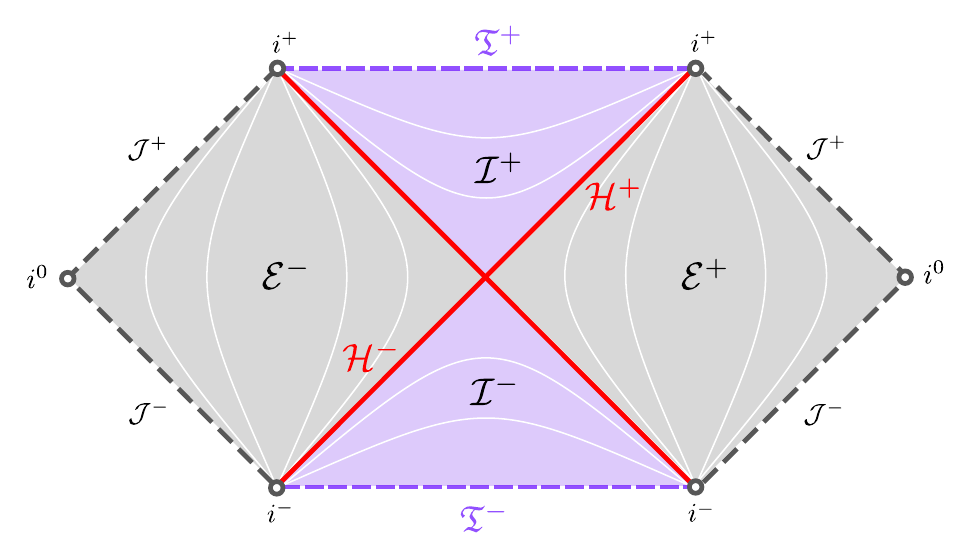}
    \caption{\justifying\footnotesize Conformal diagram of the maximally extended spacetime \cite{Alonso-Bardaji:2022ear,Alonso-Bardaji:2024tvp}. The horizon $\hor$ is drawn in red, and the novel boundaries $\mathfrak{T}$, which replace the singularity, are depicted by the dashed purple lines. Dashed gray lines represent past and future null infinities, and the gray rings correspond to timelike and spacelike infinities. Additional constant $r$ hypersurfaces are shown in white, and the trapped regions are shaded in purple.}
    \label{fig.diagram}
\end{figure}

A straightforward calculation shows that the domain of the above chart is {\red a part of the} static nontrapped region $\mathcal{E}:=\{r>r_g\}$, where the surfaces of constant $r$ are timelike. Likewise, it is valid in the homogeneous trapped domain $\mathcal{I}:=\{r_0<r<r_g\}$, where the level surfaces of $r$ are spacelike. 
As in Schwarzschild, $\hor:=\{r=r_g\}$ defines the lightlike trapping---and Killing---horizon (with area $A:=4\pi r_g^2$), which can be reached in finite proper time. In contrast, the set~$\mathfrak{T}:=\{r=r_0\}$ is attained for divergent values of the affine parameter $s$ of radial causal geodesics, even though the area of the spheres of constant time and radial coordinates remains finite there. In fact, it would achieve its infimum $\Delta:=4\pi r_0^2$. More precisely (and removing the zero-energy null geodesics lying on $\hor$), the interval $s\in\mathbb{R}$ is the preimage of $r\in(r_0,\infty)$ {\red (for the detailed computation, see Sec.~V~A of Ref.~\cite{Alonso-Bardaji:2024tvp})}. 
The singularity is thus replaced by an asymptotic non-traversable spacelike boundary of the trapped region of spacetime, where all curvature scalars converge to finite values. {\red More precisely, the curvature can be completely characterized by the Ricci scalar ($R$) and the Coulombian term of the Weyl tensor ($\Psi_2$), which read $R|_{\mathfrak{T}}=-8\Psi_2|_{\mathfrak{T}}=\frac{1}{24r_0^3}(r_g+3r_0)$ on this boundary. Note that these values are linear in $r_g$, and thus increase with the size of the black hole.}
The conformal diagram of the maximally extended spacetime is represented in Fig.~\ref{fig.diagram}, where one can see that the four sets $\mathcal{E}$, $\hor$, $\mathcal{I}$, and $\mathfrak{T}$ are composed of two disconnected regions, denoted by a $\pm$ superscript.

Let us remark that the maximal extension of the geometry comprises a bifurcate Killing horizon $\hor$ of the Killing field $t^\mu\partial_\mu$, which is completely analogous to the Schwarzschild case, and the computation of the Hawking-radiation spectrum faces no new conceptual issues. This means that on any of the asymptotically flat exterior regions $\mathcal{E}^\pm$, there exists a thermal state for the quantum fields propagating on the background spacetime~\eqref{eq.metric}. 
In other words, we find an induced spectrum of temperature 
\begin{align}\label{eq.temperature}
    T:=\frac{1}{2\pi}\sqrt{-\frac{1}{2}(\nabla_\mu t_\nu)(\nabla^\mu t^\nu)}\bigg|_{\hor} = \frac{1}{4\pi r_g}\sqrt{1-\frac{r_0}{r_g}},
\end{align}
which is proportional to the surface gravity of the horizon. But in clear contrast to the classical case, 
it attains an absolute maximum  
at $r_g=3r_0/2$ and vanishes as $r_g\to r_0$.

If we wish to further explore the thermodynamical properties of the black hole, we need to define the concept of energy. Let us consider the Misner-Sharp {\red energy},
\begin{align}\label{eq.energy}
    E:=\frac{r}{2}\big(1- \nabla_\mu r \nabla^\mu r \big)=\frac{1}{2}r_g+{r_0}\bigg(1-\frac{r_g}{r}\bigg)\bigg(1-\frac{r_0}{2r}\bigg),
\end{align}
which is a quasi-local notion of the energy contained within any sphere of radius $r$. Unlike in Schwarzschild, it is not constant; however, on the horizon, it retains the same value, namely, $E|_\hor=\frac{1}{2}r_g$ and $dE|_\hor=\frac{1}{2}d r_g$. Using the first law of thermodynamics, $dE=T dS$, we are able to determine the entropy associated with the black-hole horizon:
\begin{align}\label{eq.entropy}
    S&
    =\pi r_g^2\left(1+\frac{3r_0}{2r_g}\right)\sqrt{1-\frac{r_0}{r_g}}+\frac{3}{4}\pi r_0^2\log\frac{\sqrt{r_g}+\sqrt{r_g-r_0}}{\sqrt{r_g}-\sqrt{r_g-r_0}} +S_0 .
\end{align}
In agreement with the third law of thermodynamics, the constant \textit{minimum} entropy $S=S_0$, which we set to zero, is attained in the vanishing temperature case. It is also interesting to point out that the maximum-temperature case $r_g=3r_0/2$ is an inflection point of the entropy function.
Remarkably, this formula brings us back to loop quantum gravity: In the astrophysical regime, $r_g\gg r_0$, the entropy reads $S\approx\pi r_g^2 +\pi r_0 r_g+\frac{3}{4}\pi r_0^2\log r_g$, which coincides with full spin-network calculations~\cite{Ghosh_2014}. 

Once this connection is established, let us discuss the area operator in loop quantum gravity. According to the full quantum kinematics, it has a positive minimum eigenvalue proportional to the Barbero-Immirzi parameter ($\gamma$), which is a fundamental constant of the theory that remains undetermined. By identifying this area gap with our infimum, we find $r_0^2=\sqrt{3}\gamma$. In turn, the polymerization parameter becomes a running constant, varying over the space of solutions (like the mass does in Schwarzschild) as $\lambda^2 = r_0 / (r_g - r_0)$. 
Consequently, astrophysical black holes ($r_g \gg r_0$) have a negligible polymerization parameter while still being free of singularities.  
{\red This phase-space-dependent constant polymerization thus solves the problem of large quantum corrections in low curvature regions, which may appear in other schemes \cite{Ashtekar:2018cay}.} 
{\red Besides,} quantum gravity effects become {\red now} increasingly significant as we approach the infimum $r_g \to r_0$, where the polymerization parameter diverges {\red and marks} the boundary of validity for the model.

In the following, we prove that this infimum cannot be achieved through classical black-hole evaporation. Assuming that the quasi-static approximation remains valid, thermal emission produces stable objects with nearly zero temperature and entropy, which we shall call \textit{black-hole remnants}. 
When modeled as a perfect black body, thermal emission follows the Stefan-Boltzmann law, ${dE}/{d\tau} = {\pi^2} A T^4/{60}$ in natural units. This energy loss causes the horizon to shrink {\red as}
\begin{align}\label{eq.evaporation}
    \frac{dr_g}{d\tau}=-\frac{1}{1920\pi\,r_g^2}\bigg(1-\frac{r_0}{r_g}\bigg)^2,
\end{align}
{\red where $\tau$ is a formal time to follow the evaporation through the parameter space.}

\begin{figure}[t]
    \centering
    \begin{minipage}[b]{0.46\textwidth}
        \centering
        \includegraphics[width=\linewidth]{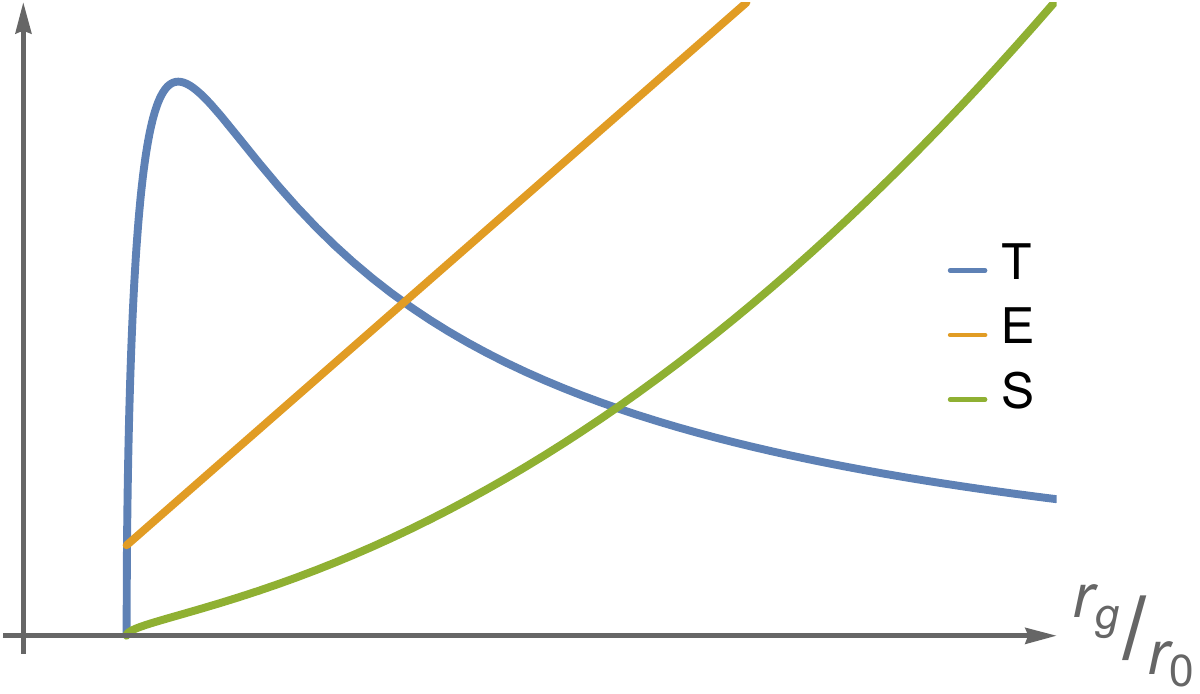}
        \caption{\justifying\footnotesize Temperature (blue), energy (orange), and entropy (green) associated with the black-hole horizon [scale $100\!:\!1\!:\!1/100$]. For large $r_g$, their behavior follows $1/r_g$, $r_g$, and $r_g^2$, respectively. The {\red zeros} of $T$ and $S$ are at $r_g=r_0$.} 
    \end{minipage}
    \hfill
    \begin{minipage}[b]{0.45\textwidth}
        \centering
         \includegraphics[width=\linewidth]{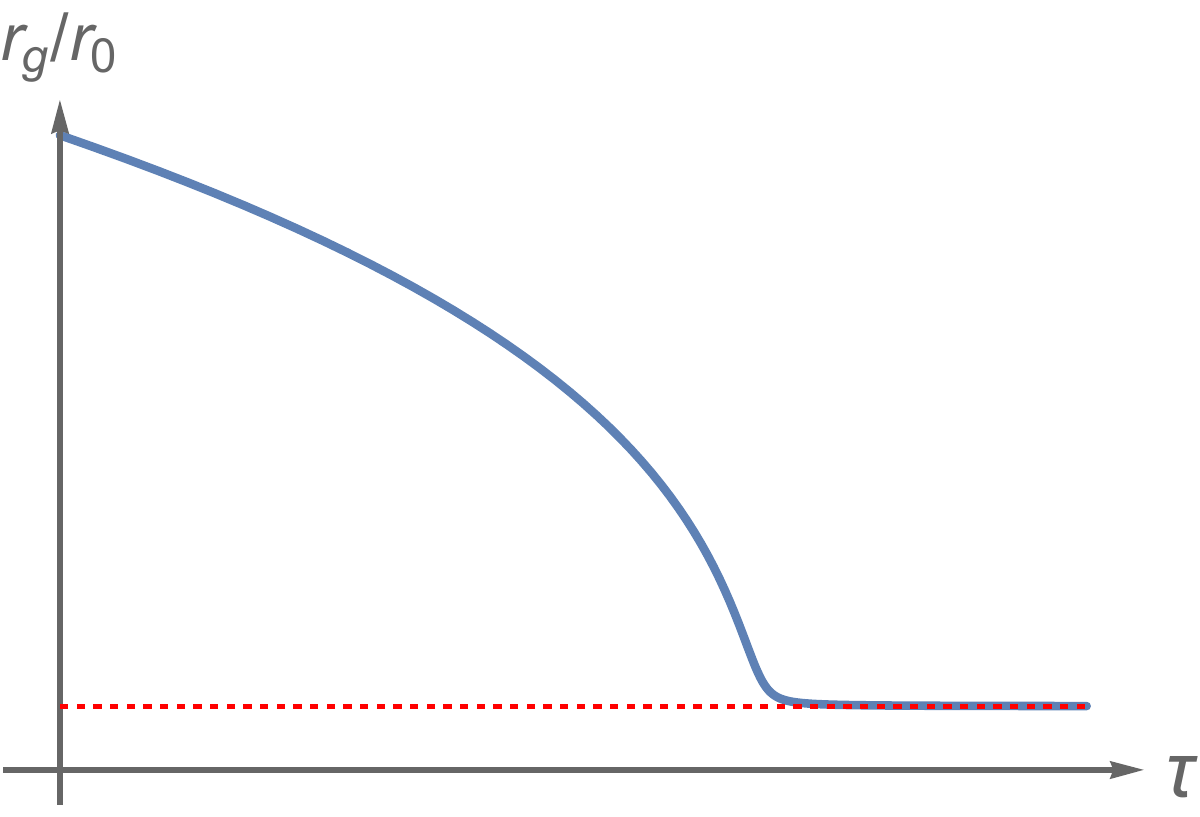}
    \caption{\justifying\footnotesize Final stages of the evaporation process (starting from $r_g=10r_0$). The curve has an inflection point at $r_g=2r_0$, and asymptotically approaches $r_g=r_0$ (dashed red line).}
    \label{fig.m(t)}
    \end{minipage}
\end{figure}

The function $r_g(\tau)$ is monotonically decreasing, with an inflection point at $r_g=2r_0$, where the evaporation begins to decelerate. Direct integration yields
\begin{align*}
   \tau(r_g)\propto\frac{R_g^3-r_g^3}{3}+{r_0}(R_g^2-r_g^2)+3r_0^2(R_g-r_g)
    +4r_0^3\log\frac{R_g-r_0}{r_g-r_0}+\frac{r_0^4}{r_g-r_0}-\frac{r_0^4}{R_g-r_0},
\end{align*}
with $R_g$ the initial size. 
For large black holes, the difference from the ``classical'' time to achieve a final gravitational radius of a few~$r_0$ scales as $r_0 R_g^2$ and is thus negligible compared to the whole process ($\tau\sim R_g^3$). As shown in Fig.~\ref{fig.m(t)}, the final remnant state requires an infinite amount of time to form. Full quantum-gravity phenomena, which lie beyond the scope of this study, might also lead to the spontaneous decay of remnants~\cite{Kazemian:2022ihc}. 
Despite all, the estimated half-life for such processes is $R_g^4$, meaning that the remnants predicted by this effective theory are metastable at worst. If their mass falls within the suitable regime, they could stand for long-lived dark-matter particles.

To conclude, let us estimate the mass of these remnants. First, we must introduce a meaningful definition of ``mass'', which we did not provide so far. Despite the existence of several inequivalent notions (recall that Einstein's equations are not satisfied), we adopt the asymptotic value $M:=\frac{1}{2}r_g+r_0$ of the Misner-Sharp energy \eqref{eq.energy}, which is a geometric invariant and coincides with the ADM mass. The mass of black-hole remnants (those with $r_g=r_0$), as measured by asymptotic observers, is thus $M_0:=\frac{3}{2}r_0$. Exploiting the relation between $r_0$ and loop quantum gravity found above, we obtain
\begin{align}\label{eq.massremnant}
    M_{0}=\sqrt[4]{\frac{243}{16}\gamma^2}\,m_P \approx0.9621m_P=20.94\mu \mathrm{g},
\end{align}
where we have used the widely accepted value $\gamma=0.2375$ to provide a numerical estimate~\cite{Meissner:2004ju}, and have reintroduced units for clarity. 
Hence, the effective theory predicts not only the existence but also the formation of certain objects at the natural scale of quantum gravity~$m_P$. These ``particles'' are expected to interact solely gravitationally, making them an intriguing candidate for dark matter \cite{Bianchi:2018mml, Amadei:2021aqd, Rovelli:2024sjl}. Their direct observation---as suggested, for example, in Ref.~\cite{Perez:2023tld}---could provide an experimental measure of the Barbero-Immirzi parameter. For instance, if the remnant was exactly one Planck mass, we would find $\gamma=0.2566$, lying within the expected range $\log(2)\leq\pi\gamma\leq\log(3)$ \cite{Domagala:2004jt}.

{\red In Ref.~\cite{Alonso-Bardaji:2024tvp}, we show that the geometry~\eqref{eq.metric} is, in fact, the vacuum solution corresponding to the nonsingular collapse of a spherical dust cloud in effective loop quantum gravity. As such, the ``pure’’ vacuum solution would only be realized in the remnant state, which can indeed be interpreted as a quantum excitation of the gravitational field.} 
In complete agreement with the third law of black-hole thermodynamics, this final state---characterized by zero temperature and entropy---cannot be achieved in finite time. 
The theory also allows us to determine the mass of these {\red black-hole} remnants, predicting 20.94$\mu$g for the value $\gamma=0.2375$ of the Barbero-Immirzi parameter. If the effective theory remains valid at these scales, it provides an interesting framework to study a promising long-lived dark-matter candidate. 
As a final remark, let us point out that the \textit{limit} of vanishing corrections ($r_0\to0$) is well defined and reproduces the classical expressions and results. In particular, the novel boundary $\mathfrak{T}$ emerges as the Schwarzschild singularity (see Fig.~\ref{fig.diagram}) and causes black holes to evaporate in finite time.

\textbf{Acknowledgments}. The author thanks David Brizuela and Alejandro Perez for interesting comments. This work was made possible through the Grant PID2021-123226NB-I00 (funded by MCIN/AEI/10.13039/501100011033 and by “ERDF A way of making Europe”) and the support of the ID\#~62312 grant from the John Templeton Foundation, as part of the project \href{https://www.templeton.org/grant/the-quantum-information-structure-of-spacetime-qiss-second-phase}{``The Quantum Information Structure of Spacetime'' (QISS)}. The opinions expressed in this work are those of the authors and do not necessarily reflect the views of the John Templeton Foundation.

\providecommand{\noopsort}[1]{}\providecommand{\singleletter}[1]{#1}%

\end{document}